\documentclass[3p,times,procedia]{elsarticle}
\usepackage{nupha_ecrc}

\volume{00}

\firstpage{1}

\journalname{Nuclear Physics A}

\runauth{}

\jid{nupha}

\jnltitlelogo{Nuclear Physics A}

\usepackage{amssymb}

\usepackage[figuresright]{rotating}

\usepackage{amsmath}
\usepackage{slashed}
\usepackage{amssymb}
\usepackage{mathtools}
\usepackage{graphicx}
\usepackage{subfigure}
\usepackage{setspace}
\usepackage{sidecap}
\usepackage{color}
\usepackage{hyperref}
\usepackage{tabu} 
\usepackage{hyperref}
\usepackage[export]{adjustbox}
\usepackage{tensor}

\begin{document}

\newcommand{\lb}{\Big{\lbrack}}
\newcommand{\rb}{\Big{\rbrack}}
\newcommand{\lp}{\Big{(}}
\newcommand{\rp}{\Big{)}}
\newcommand{\lbc}{\Big{\lbrace}}
\newcommand{\rbc}{\Big{\rbrace}}
\newcommand{\nn}{\nonumber}
\newcommand{\Bvert}{\Big{\vert}}
\newcommand{\Rangle}{\Big{\rangle}}
\newcommand{\Langle}{\Big{\langle}}
\newcommand{\bm}[1]{{\boldsymbol{\mathrm{#1}}}}
\newcommand{\bmax}{b_{\text{max}}}
\newcommand{\Q}[4]{ {}^{#1} #2 ^{[#4]}_{#3} }
\newcommand{\ve}{\lambda}

\begin{frontmatter}



\dochead{XXVIIIth International Conference on Ultrarelativistic Nucleus-Nucleus Collisions\\ (Quark Matter 2019)}

\title{An Effective Theory of Quarkonia in QCD Matter}


\author[yiannis]{Yiannis Makris}
\author[ivan]{Ivan Vitev}

\address[yiannis]{INFN Sezione di Pavia, via Bassi 6, I-27100 Pavia, Italy}
\address[ivan]{Los Alamos National Laboratory, Theoretical Division, Group T-2, MS B283, Los Alamos, NM 87545, USA}

\begin{abstract}
The problem of quarkonium production in heavy ion collisions presents a set of unique theoretical challenges -- from the relevant production mechanism of $J/\psi$ and $\Upsilon$
to the relative significance of distinct cold and hot nuclear matter effects in the observed attenuation of quarkonia. In 
these proceedings we summarize recent work on the generalization of non-relativistic Quantum Chromodynamics (NRQCD)  to include
off-shell gluon (Glauber/Coulomb) interactions in strongly interacting matter. This new effective theory  provides for the first time a universal microscopic description of the in-medium interaction of heavy quarkonia, consistently applicable to a range of phases such as cold nuclear matter, dense hadron gas, and quark-gluon plasma. It is an important step forward in understanding the common trends in proton-nucleus and nucleus-nucleus data on quarkonium suppression. We derive explicitly the  leading and sub-leading interaction terms in the Lagrangian and show the connection of the leading result to existing phenomenology.
\end{abstract}


\end{frontmatter}


\section{Introduction}
\label{sec:intro}

Calculations of  quarkonium production encounter hierarchies of momentum and mass scales, which allows for the  construction of effective filed theories (EFTs) that can reduce theoretical uncertainties and improve computational accuracy. The established theory of the $J/\psi$ and  $\Upsilon$ families' production and decays is non-relativistic QCD (NRQCD)~\cite{Bodwin:1994jh}.  More recently matching onto the effective theory and the renormalization group equations for NRQCD were investigated, leading to velocity renormalization group NRQCD (vNRQCD)~\cite{Luke:1999kz} and potential NRQCD (pNRQCD)~\cite{Pineda:1998kj}.

Quarkonium production in heavy ion collisions remains a multiscale problem. In spite of recent phenomenological advances, a fully coherent theoretical picture of quarkonium production at the Relativistic Heavy Ion Collider (RHIC) and the Large Hadron Collider (LHC)  has not yet emerged.  In proton-nucleus (p+A) collisions, where QGP is  less likely  to be formed, attenuation of the $J/\psi$ and  $\Upsilon$ states  similar to the one  in nucleus-nucleus (A+A) collisions is seen, though the magnitude is smaller.  The  CMS collaboration found evidence    that
 even in high multiplicity  proton-proton (p+p) collisions the $\Upsilon(2S)$ disappearance as a  function of the hadronic activity (${\rm N}_{\rm tracks}$) in the  event.  Differential  $\psi', \, \chi_c$ and $\Upsilon$ suppression was also  experimentally found at RHIC~\cite{Adare:2013ezl,Adamczyk:2013poh} in d+Au reactions.  There is an opportunity to  further develop {\em microscopic} QCD approaches that describe this quarkonium physics in nuclear matter and that will facilitate  the quantitative determination  of the transport properties of the QGP and the hadron gas. With this motivation, we summarize the development of an EFT for $J/\psi$ and $\Upsilon$ production in reactions with heavy nuclei~\cite{Makris:2019ttx}.

\section{Challenges to energy loss phenomenology}
\label{sec:e-loss}

Before we proceed to  the formulation of a generic effective theory of quarkonium production in matter, we have to explore whether medium-induced radiative processes might contribute significantly to the  modification of quarkonium cross sections in reactions with nuclei. It was suggested~\cite{Spousta:2016agr}  that such effects can reduce the cross section of high transverse momentum $J/\psi$ production at the LHC~\cite{Khachatryan:2016ypw,Aaboud:2018quy}. In order for energy loss processes to contribute significantly to the modification of quarkonium cross sections in QCD matter, quarkonium production must be expressed as fragmentation of partons into the various $J/\psi$ and $\Upsilon$ states.  This is possible within the leading power factorization approach to QCD.

\begin{figure}[h!]
\begin{center}
\includegraphics[width =  0.46\textwidth]{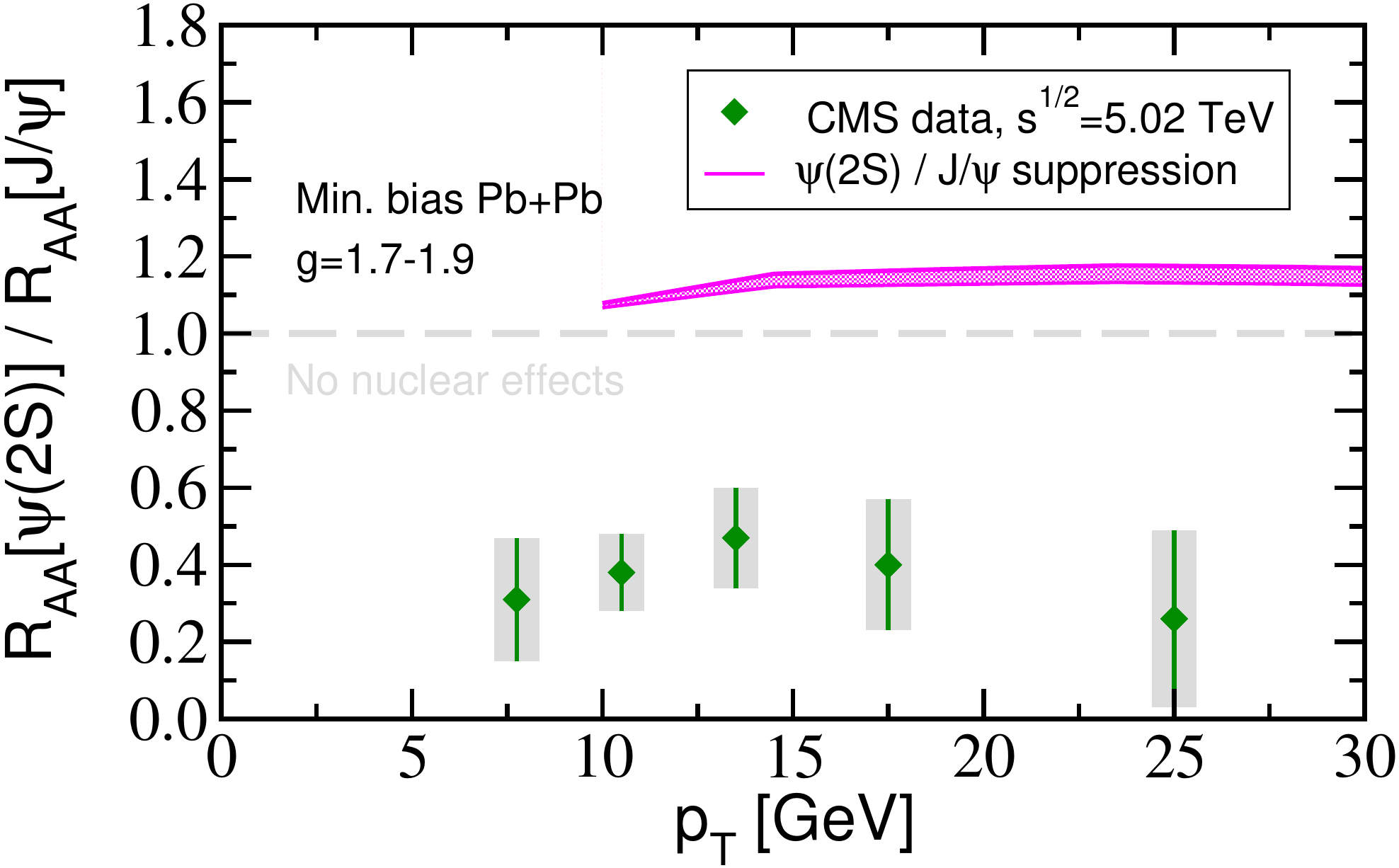}  \  \
 \includegraphics[width =  0.49\textwidth]{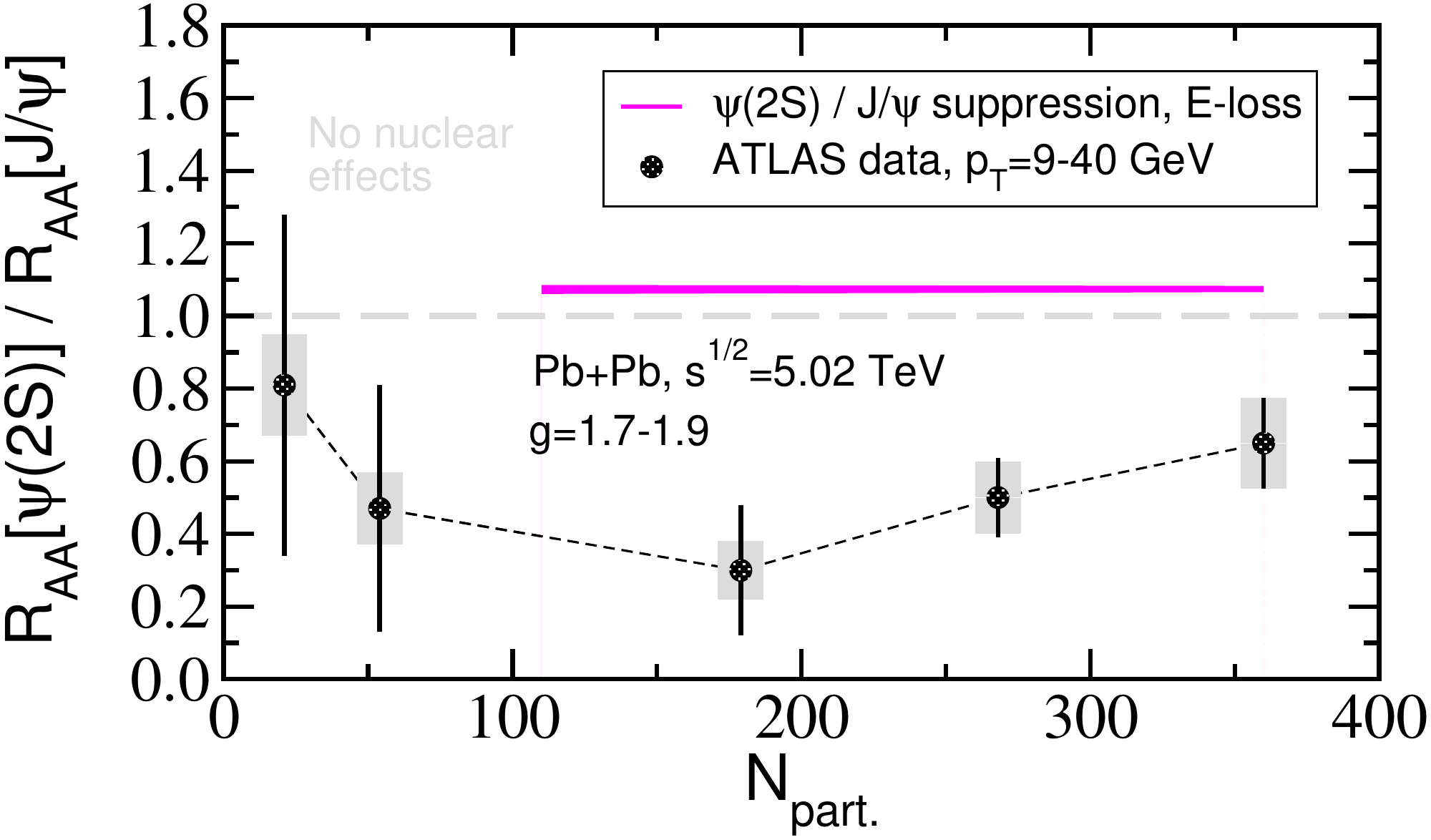} 
 \end{center}
  \caption{ The double ratio of $\psi(2S)$ to $J/\psi$ suppression (purple bands) as a measure of the relative significance of QCD matter effects on
  ground and excited states  is compared to energy loss model calculations. Left panel:  comparison  between theory and CMS data~\cite{Khachatryan:2016ypw}  as a function of  transverse momentum $p_T$ for minimum bias collisions.  Right panel:   comparison between theory and ATLAS data~\cite{Aaboud:2018quy} as a function of centrality  integrated in the  $p_T$
  region of 9-40~GeV.  } 
  \label{fig:2Sto1Sptcent}
\end{figure}

To calculate the baseline $J/\psi$ and $\psi(2S)$ production we use an extraction of the non-perturbative long-distance matrix elements (LDMEs) consistent with LP factorization, as given  by Ref.~\cite{Bodwin:2015iua}.  The energy loss evaluation is carried out in the soft gluon emission limit of the full in-medium splitting kernels~\cite{Ovanesyan:2011kn,Kang:2016ofv,Sievert:2019cwq}.  We find that the energy loss approach overpredicts the suppression of $J/\psi$ in applicable   $p_T > 10$~GeV range by a factor of 2 to 3 in both minimum bias and central collisions.  The most important discrepancy, however, is in
the relative medium-induced suppression of $\psi(2S)$ to $J/\psi$ in matter as shown in Figure~\ref{fig:2Sto1Sptcent}.  The purple bands correspond  to variation of the
coupling between the parton and the medium of $g=1.7-1.9$.  The left panel of  Figure~\ref{fig:2Sto1Sptcent} shows the double nuclear modification ratio as a function of $p_T$ compared to CMS data~\cite{Khachatryan:2016ypw}.  The right panel of  Figure~\ref{fig:2Sto1Sptcent} shows the same ratio as a function of the number of participants $N_{\rm part.}$ and the transverse momenta are integrated in the range of 9-40~GeV.  The energy loss model predicts smaller suppression for the  $\psi(2S)$ state when compared to $J/\psi$  and $R_{AA}[\psi(2S)] / R_{AA}[J/\psi] \sim 1.1$.  The experimental results show that the suppression of the weakly bound   $\psi(2S)$ is 2 to 3 times larger than that of $J/\psi$ and the tension between data and the calculations  is inherent to the  theoretical model. 

\section{Non-relativistic QCD with Glauber gluons}
\label{sec:NRQCDG}

When an energetic particle propagates in matter, the interaction with the quasiparticles of the medium is  typically mediated  by off-shell gluon exchanges. When the momentum exchanges correspond to forward $t$-channel scattering with momentum scaling $q^{\mu}_G \sim (\lambda^{2},\lambda^{1},\lambda^{1},\lambda^{2}) $ we call such modes  Glauber gluons.  The new EFT which incorporates Glauber interactions, we refer to as NRQCD$_\text{G}$. When the  sources  of interaction do not have large (three-)momentum component~\cite{Ovanesyan:2011xy},  the exchange gluon field's momentum can scale as $q^{\mu}_C \sim (\lambda^{2},\lambda^{1},\lambda^{1},\lambda^{1})$  and we call them Coulomb gluons. As a next step we work out the scaling of the Glauber and Coulomb fields $A_{G/C}^\mu$.  It depends on the source of in-medium interactions -- collinear, static, or soft -- and is summarized in Table~\ref{tb:scaling-A}.

\begin{table}[h!]
  \renewcommand{\arraystretch}{1.4}
  \begin{center}
    \begin{tabular}{|r|c|c|c|}
      \hline
      Source & Collinear      & Static       & Soft    \\ \hline \hline
      $\;\;A_C^{\mu} \sim $  &                 n.a.                       &$(\lambda^1,\lambda^2,\lambda^2,\lambda^2)$&$(\lambda^1,\lambda^1,\lambda^1,\lambda^1)$   \\ \hline
      $\;\;A_G^{\mu} \sim $  & $(\lambda^2,\lambda^3,\lambda^3,\lambda^2)$ &                 n.a.                     &               n.a.                           \\ \hline
    \end{tabular}
    \caption{The Glauber/Coulomb filed scaling for different sources of interaction in matter.}
    \label{tb:scaling-A}
  \end{center}
\end{table}

We can now construct the Lagrangian of NRQCD$_{\rm G}$ by adding to the vNRQCD Lagrangian the  terms that include the interactions with medium sources through virtual Glauber/Coulomb gluons exchanges. It takes the form:
\begin{equation}
  \mathcal{L}_{\text{NRQCD}_{\rm G}} = \mathcal{L}_{\text{vNRQCD}} + \mathcal{L}_{Q-G/C} (\psi,A_{G/C}^{\mu,a}) + \mathcal{L}_{\bar{Q}-G/C} (\chi,A_{G/C}^{\mu,a})\;,
\end{equation}
where in the background field method the effective field $A_{G/C}^{\mu,a}$ incorporates the information about the sources. To extract the form and perform the power-counting of the terms in  $\mathcal{L}_{Q-G/C} (\psi,A_{G/C}^{\mu,a})$ we follow three different approaches: 
\begin{enumerate}
\item Perform a shift in the gluon field in the NRQCD Lagrangian ($A^{\mu}_{us} \to A^{\mu}_{us} +A^{\mu}_{G/C}$)  and then perform the power-counting established in Table~\ref{tb:scaling-A} to keep the leading contributions. This approach is also known as the background field method.
\item A hybrid method, where from the full QCD diagrams for single effective Glauber/Coulomb gluon insertion, and after performing the corresponding power-counting, one can read the Feynman rules for the relevant interactions. 
\item A matching method  where we expand in the power-counting parameter, $\lambda$, the full QCD diagrams describing the interactions of an incoming heavy quark and a light quark or a gluon. To get the NRQCD$_{\rm G}$ Lagrangian, we then keep the leading and subleading contributions and focus on the dominant contributions in forward scattering limit.  In contrast to the  hybrid method, here we also derive the tree level expressions of the effective fields in terms of the QCD ingredients.  
\end{enumerate}

We are now ready to write the leading and subleading correction to the NRQCD$_{\rm G}$ Lagrangian in the heavy quark sector from virtual (Glauber/Coulomb) gluon insertions, i.e. $\mathcal{L}_{Q-G}$, :
\begin{equation}
  \label{eq:L0-NR}
  \mathcal{L}_{Q-G/C}^{(0)} (\psi,A_{G/C}^{\mu,a})  = \sum_{{\bf p},{\bf q}_T}\psi^{\dag}_{{\bf p}+{\bf q}_T} \lp - g A^{0}_{G/C} \rp \psi_{{\bf p}}\;\; (collinear/static/soft)\; , 
\end{equation}
and 
\begin{align}
  \label{eq:L1-NR}
  \mathcal{L}_{Q-G}^{(1)} (\psi,A_{G}^{\mu,a}) & =  g\sum_{{\bf p},{\bf q}_T} \psi^{\dag}_{{\bf p} +{\bf q}_T} \lp\frac{ 2  A_{G}^{\bf n} ({\bf n} \cdot {\bf \mathcal{P}}) - i \lb ( {\bf \mathcal{P}}_{\perp} \times {\bf n}) A^{{\bf n}}_{G}  \rb \cdot \bm \sigma }{2m} \rp \psi_{{\bf p}}\;\; (collinear)\; , \nn\\
  \mathcal{L}_{Q-C}^{(1)} (\psi,A_{C}^{\mu,a})  &= 0\;\; (static)\;,\nn\\
  \mathcal{L}_{Q-C}^{(1)} (\psi,A_{C}^{\mu,a}) & =  g\sum_{{\bf p},{\bf q}_T} \psi^{\dag}_{{\bf p} +{\bf q}_T} \lp\frac{ 2  {\bf A}_{C} \cdot {\bf \mathcal{P}} + [{\bf \mathcal{P}} \cdot  {\bf A}_{C} ] - i \lb  {\bf \mathcal{P}} \times {\bf A}_{C} \rb \cdot \bm \sigma }{2m} \rp \psi_{{\bf p}}\;\; (soft)\;,
\end{align}
where we use squared brackets in order to denote the region in which the label momentum operator, $\mathcal{P}^{\mu}$, acts. 
We find that to lowest order (LO) the modification in  the  leading Lagrangian, $\mathcal{L}_{Q-G/C}^{(0)}$, is independent of the 
momentum scaling of the quasiparticles of the QCD medium. In contrast, at next-to-leading order (NLO) the distinction is manifested in the subleading Lagrangian, $\mathcal{L}_{Q-G/C}^{(1)}$.

\section{Conclusions}
\label{sec:onclude}

Motivated by the commonality in  the $J/\psi$ and $\Upsilon$ suppression trends in high-multiplicity p+p, p+A, and A+A reactions and the 
challenges that quarkonia present to traditional energy loss phenomenology, we extended non-relativistic QCD to include Galuber/Coulomb gluon interactions in nuclear matter~\cite{Makris:2019ttx}.  We derived the NRQCD$_{\rm G}$ leading and sub-leading Lagrangians for a single virtual gluon exchange using three different approaches: i) a background field method, ii) a matching (with QCD) procedure, and iii) a hybrid method. All three methods give the same Lagrangian as a non-trivial check of our results. In the future we can extend this work to  double virtual gluon interactions  with minimal effort in the background field method.

In Ref.~\cite{Makris:2019ttx} we focused on the formal aspects of  of NRQCD$_{\rm G}$,  leaving phenomenological applications  for future investigation.  Still, we note that the collisional dissociation rates of quarkonia employed in~\cite{Sharma:2012dy,Aronson:2017ymv} correspond to the lowest order Lagangian of the new theory with an averaging to a color neutral final state. Using this EFT, it will  be informative  to study the modification of the heavy quark-antiquark potential due to medium interactions. Furthermore,  interactions with the medium could induce radial excitations on a short time scale,  which will likely cause transitions from one quarkonium state to another.  Medium-induced transitions from and to exited states might also modify the observed relative suppression rates or ground and excited states.  











\end{document}